# Constraint evolution in first-order viscous relativistic fluids

Delfina Fantini[1*] and Marcelo E. Rubio[2,3†]

[1] *Facultad de Matemática, Astronomía, Física y Computación, Universidad Nacional de Córdoba, Av. Medina Allende s/n, X5000 Córdoba, Argentina*
[2] *Gran Sasso Science Institute (GSSI), Viale Francesco Crispi 7, I-67100 L'Aquila, Italy*
[3] *INFN, Laboratori Nazionali del Gran Sasso, I-67100 Assergi, Italy*

Relativistic hydrodynamics provides a solid framework for evolving matter and energy in a wide variety of phenomena. Nevertheless, the inclusion of dissipative effects in realistic scenarios through causal, stable, and well-posed theories still constitutes an open problem. In this paper, we study the evolution of the algebraic and differential constraints stemmed from the first-order reduction proposed by Bemfica, Disconzi, Noronha and Kovtun (BDNK), for proving the local well-posedness of conformally-invariant viscous fluids in Sobolev spaces. First, we show analytically that the whole set of constraints satisfies a homogeneous, strongly-hyperbolic system of equations, ensuring a correct propagation as a consequence of the fluid equations. Motivated by this result, we explore their numerical stability by performing simulations of the BDNK reduction restricted to plane-symmetric configurations, in flat spacetime. We report on different initial data sets initially satisfying the constraints, and whose evolution leads to stable configurations. This result suggests that the proposed reduction by BDNK is suitable for numerical evolutions, keeping the constraints accurate under small numerical errors.

## I. INTRODUCTION

The multi-messenger detection of gravitational waves from binary neutron-star mergers [1, 2], together with their electromagnetic counterparts through extraordinarily violent short gamma-ray bursts [3, 4], over again marked down the significance of getting a consistent modeling of matter viscous effects in strong gravity. Such extreme phenomena involving highly-nonlinear dynamics are essential for placing constraints on gravity in the strong field regime [5–8], since the interaction of dense matter would otherwise not be assessable. Moreover, the advent of third-generation detectors will shed light on many aspects on the physics of neutron stars and nuclear matter [9, 10]. Among them, they will help to unveil how "stiff" the interior of a star can be; when do phase transitions to quark matter take place, what is the origin of bulk viscosity as a micro-physical transport effect; and how the nucleosynthesis of heavy elements is affected by neutrino transport [11–15].

Relativistic hydrodynamics is perhaps the most suitable workhorse for describing the macroscopic behavior of matter in extreme conditions, with a scale range running from heavy-ion collisions [16] and high-energy astrophysics [17] to the very early Universe [18]. However, understood as an effective theory in the long-wavelength limit of any system of many particles, hydrodynamics still constitutes a broadly open field of study, despite being one of the oldest in classical physics.

The mathematical formulation of any theory aiming to describe fluid dynamics is based on conservation laws, provided an equation of state which accounts for microscopic effects of the fluid elements [19–21]. The simplest approach is the ideal fluid, which describes a macroscopic, homogeneous and isotropic system in thermal equilibrium, and whose dynamics is governed by the Euler's equations. While this formulation results effective as a first approximation to the phenomenology of fluids, it is not entirely accurate for modeling scenarios involving energy transport through heat and viscosity. This necessarily motivates the inclusion of various viscous effects, according to the scales of the problem at hand. A classic example is constituted by the Navier-Stokes equations, which allow a more accurate evolution of macroscopic phenomena, and where the ideal approximation renders the description of the system inconsistent [22, 23].

The Navier-Stokes equations are effective in a non-relativistic framework. Indeed, they result insufficient for the study of astrophysical processes where: (i) either the characteristic velocities are close to the speed of light –such as the modeling of astrophysical jets [24–26]–; or (ii) where gravitational effects become very intense, such as the dynamics of matter surrounding a supermassive black hole, the stability of accretion disks [27], or the evolution of neutron-star mergers [17, 28], among other phenomena. It is in these regimes that a fully relativistic description of viscous fluids becomes essential. However, including dissipative effects in the context of relativistic hydrodynamics leads to mathematical inconsistencies in the evolution equations [29–31], still constituting an open problem in the field.

The first covariant formulations attempting to account for viscous effects in the description of relativistic fluids date back to Eckart, in 1940 [32], and Landau-Lifshitz, in 1953 [22]. Although both of them reached a fully-covariant formulation by means of constitutive relations depending on first gradients of the thermodynamic variables, they failed to be effective due to generic linear instabilities, thus predicting unbounded fluid modes in the high-frequency limit. This seminal study was carefully carried out by Hiscock and Lindblom in 1984 [33], suggesting that first-order theories are unsuitable for physical predictions through numerical simulations. To address these shortcomings, Müller, Israel and Stewart proposed going beyond first-order theories, and considered instead second-order contributions for modeling viscous effects [34]. By introducing additional evolution equations for the dissipative fields, the authors were able to restore the causality and stability of the theory, allowing to perform the first numerical simulations on viscous fluids in stronger regimes. The nonlinear regime of the theory was further understood in [35]. However, their approach relied on a truncated expansion of non-equilibrium effects, introducing additional complexity and new parameters in the equations [21, 29]. This also required sophisticated numerical techniques to solve them efficiently, giving rise to the "fixing-the-equations" approach [36]. As a consequence, all this motivated the development of new formulations for viscous fluids, among which are the "divergence-type" families of theories [37, 38]; yielding stability and causality only if second-order contributions are considered [39].

* delfina.fantini@mi.unc.edu.ar
† marcelo.rubio@gssi.it

More recently, Bemfica, Disconzi, Noronha and Kovtun (BDNK) suggested an interesting alternative framework for treating viscous fluids which, even assuming first-order dissipative contributions, it surprisingly gives rise to causal and linearly stable theories, provided some restrictions on the parameter space [40–43]. This approach opened a window for retrieving first-order models, actually constituting a promising candidate for evolving viscous fluid systems, particularly in the context of numerical relativity simulations [44–48]. But in order to achieve stable evolutions, it is essential to require the theory to constitute a *well-posed* initial-value problem.

Well-posedness is a local concept which means that, for any given initial data set in certain functional space, there must exist (locally) a unique solution. Additionally, the map between the space of initial data and the space of solutions must be continuous, with respect to the topology considered on each functional space [49]. For physically-motivated systems of quasi-linear first-order equations, it is usual to look for well-posedness in Sobolev spaces [50–52]. If the system is not equipped with differential constraints, the standard way to achieve this goal is by studying the algebraic properties of the *principal part* of the equations (i.e., the terms with highest order in derivatives). The latter contains all the information about the propagation speeds of the theory, and the well-posedness is elucidated through the concept of *hyperbolicity*: a set of algebraic conditions the principal part must satisfy [53]. In particular, a system whose principal part is diagonalizable with real eigenvalues is referred as to be *strongly-hyperbolic* [54–56]. It can be shown that, for first-order quasi-linear partial differential equations, strong-hyperbolicity is *equivalent* to local well-posedness in Sobolev spaces [51, 57, 58].

Now, if the system admits differential or algebraic constraints (i.e., equations without time derivatives that must be satisfied along evolution), the hyperbolicity criteria can be addressed in an similar way, provided the correct propagation of the constraints. The latter means that, if the constraints are initially satisfied, then, as a consequence of the evolution equations, they must be also satisfied for further times. Nevertheless, this is not the only way to prove the local well-posedness. Another possibility, which *does not* require the inspection of the constraints, is by means of the *Schauder method* [59], based on applying the Cauchy-Kovalévskaya (CK) theorem for the existence of a solution satisfying non-characteristic initial data, and proving their uniqueness from uniform *a priori* estimates for the difference of solutions. Although this path constitutes a classical and standard tool in Functional Analysis, the strong hyperbolicity of general covariant systems with differential constraints is currently an active open problem, and we refer the reader to Refs. [60–62], for more details.

Local well-posedness of BDNK theory was first proved in a weaker sense, for which existence and uniqueness was guaranteed in certain Gevrey spaces. This was achieved by means of the *Leray* hyperbolicity criteria [63, 64], which consists of studying specific properties of the characteristic polynomial associated to the principal part. Shortly after, the authors considered the general problem of well-posedness in Sobolev spaces. In one of their works, they restricted the analysis to conformally-invariant fluids in flat space [65], suggesting a first-order reduction of the evolution equations, which required the introduction of differential constraints. The authors proved that the principal part of such reduction is diagonalizable with real eigenvalues, and derived energy estimates for analytic solutions, concluding that the theory is locally well-posed. More recently, global well-posedness of conformal BDNK system was also proved, in the small initial-data regime, and from a quasi-linear second-order formulation [66].

As a consequence of this result, the constraints needed for their reduction should display a correct a propagation. Nevertheless, the inspection of the full system of constraints has not been addressed in the literature, to the best of our knowledge. In this work, we perform an analytic study and a numerical exploration of the behavior of the full set of constraints emerging from the BDNK reduction displayed in the work [65], whose correct propagation is fundamental for guaranteeing the stability of numerical simulations.

The structure of the paper is the following. In Section II, we give a brief overview of the fundamental aspects of BDNK theory for conformally-invariant fluids, constitutive relations, stability and causality conditions. In Section III we derive a homogeneous, first-order quasilinear system for the full set of constraints, and prove it is strongly-hyperbolic, assuring correct propagation. Motivated by these results, in Section IV we present the numerical setup that has been implemented for simulating the BDNK conformal theory. Section V is committed to showing the numerical results, reporting on the evolu-tion of the constraints for smooth configurations, as well as displaying convergence and validation tests of our numerical scheme. Finally, we leave Section VI for general conclusions and future perspectives.

Throughout this work, we will work in Minkowski spacetime, and consider the $(-,+,+,+)$ signature convention for the spacetime metric. We will use geometric units such that $c = G = k_B = 1$, where $c$ is the speed of light in vacuum, $G$ is Newton's constant in four spacetime dimensions and $k_B$ is Boltzmann's constant.

## II. BDNK THEORY FOR CONFORMAL VISCOUS FLUIDS

In this section, we present a brief summary of the theory recently proposed by Bemfica, Disconzi, Noronha, and Kovtun (BDNK) for the description of viscous relativistic fluids. In particular, we restrict our analysis to the conformally-invariant version of the theory [65], as it constitutes the main purpose of this work. For a more in-depth study of BDNK theories, we refer the reader to the Refs. [29, 40].

As it is well-known, conformally-invariant descriptions of fluid systems are quite relevant in Particle Physics [67–71] and Cosmology [72–76]. The equations of motion governing such systems satisfy the property of remaining invariant under a smooth rescaling of the spacetime metric; i.e., under the transformation

$$g_{ab} \to \Omega^2 \, g_{ab} \, ,$$

where $\Omega$ is a smooth scalar field such that $\Omega > 0$ and $\nabla_a \Omega \neq 0$, being $\nabla$ the connection compatible with $g_{ab}$ [77]. Imposing this condition to the local conservation of the energy-momentum tensor, one finds that the latter must be traceless, that is

$$g_{ab} T^{ab} = 0 \, .$$

The above restriction implies that there is no intrinsic length scale associated to the theory, thus becoming scale-invariant. Indeed, this is the reason why conformal theories are relevant in Particle Physics, particularly in the dynamics of ultra-relativistic fluids at very high temperatures: in such scenarios, the rest energy (or mass) of the particles becomes negligible with respect to the kinetic and thermal contributions. Under this regime, ultra-relativistic fluid systems are usually referred as “conformal” fluids [39, 78, 79].

For a relativistic conformal fluid with four-velocity $u^a$, BDNK proposed the following energy-momentum tensor [65]:

$$T_{ab} = (\epsilon + A) \left( u_a u_b + \frac{1}{3} \Pi_{ab} \right) + 2 u_{(a} Q_{b)} - \eta \sigma_{ab} \, , \quad (1)$$

where

$$A = 3\chi \left( \frac{1}{\theta} u^c \nabla_c \theta + \frac{1}{3} \nabla_c u^c \right) , \quad (2)$$

$$Q_a = \lambda \left( \frac{1}{\theta} \Pi_a{}^c \nabla_c \theta + u^c \nabla_c u_a \right) , \quad (3)$$

$$\sigma_{ab} = \Pi_a{}^c \nabla_c u_b + \Pi_b{}^c \nabla_c u_a - \frac{2}{3} \Pi_{ab} \nabla_c u^c \, , \quad (4)$$

and the parenthesis in the sub-indices denotes tensor symmetrization. Here, $\epsilon$ is the ideal-fluid energy density, which satisfies a *pure-radiation* equation of state, namely

$$\epsilon = \epsilon_0 \, \theta^4 \, ,$$

where $\theta > 0$ the temperature field, and $\epsilon_0 > 0$ a constant. Also, $\eta$, $\chi$ and $\lambda$ are transport coefficients which, as a consequence of the conformal symmetry, they depend on $\theta^3$; i.e.,

$$\eta = \eta_0 \, \theta^3 \, ; \quad \lambda = \lambda_0 \, \theta^3 \, ; \quad \chi = \chi_0 \, \theta^3 \, ,$$

for some constants $\eta_0$, $\lambda_0$ and $\chi_0$. Finally, the symbol $\Pi^a{}_b := \delta^a{}_b + u^a u_b$ is the projector operator onto the space orthogonal to $u^a$, which satisfies the normalization condition $g_{ab} u^a u^b = -1$.

The equations of motion are obtained from the local conservation of $T^{ab}$ given in (1), namely

$$\nabla_a T^{ab} = 0, \quad (5)$$

supplemented by the unit-norm condition for $u^a$.

System (5) constitutes a set of four second-order differential equations for the four-velocity $u^a$ and the temperature $\theta$. To study the well-posedness of the corresponding initial-value problem, the authors proposed a first-order reduction of system (5) by introducing a new set of auxiliary variables, defined as derivatives of the four-velocity; that is,

$$S_a{}^b = \Pi_a{}^c \nabla_c u^b, \qquad S^a = u^c \nabla_c u^a. \quad (6)$$

From them, they obtained an augmented system, which reads

$$u^c \nabla_c A + \nabla_c Q^c + r_1 = 0 \, , \quad (7)$$

$$\Pi^{ac} \nabla_c A + 3 u^c \nabla_c Q^a + B_d{}^{ace} \nabla_e S_c{}^d + r_2 = 0 \, , \quad (8)$$

$$-\frac{1}{\chi} \Pi^{ac} \nabla_c A + \frac{3}{\lambda} u^c \nabla_c Q^a - 3 u^c \nabla_c S^a + \Pi^{ac} \nabla_c S_d{}^d + r_3 = 0 \, , \quad (9)$$

$$u^c \nabla_c S_a{}^b - \Pi_a{}^d \nabla_d S^b + r_4 = 0 \, , \quad (10)$$

$$\frac{1}{\theta} u^c \nabla_c \theta + r_5 = 0 \, , \quad (11)$$

$$u^c \nabla_c u^a + r_6 = 0 \, , \quad (12)$$

where

$$B_d{}^{ace} = -3\eta \left( \delta_d{}^a \Pi^{ce} + \delta_d{}^e \Pi^{ac} - \frac{2}{3} \delta_d{}^c \Pi^{ae} \right) ,$$

and $r_i\,,\ i=1,\cdots,6$ are smooth algebraic functions of $A$, $Q_a$, $S_a$, ${S_a}^b$, $\theta$, and $u^a$, but *not* of their derivatives. The explicit form of the functions $r_i$ were calculated and are displayed in Appendix A, as will be needed later on. Eqs. (7) and (8) are the projections of Eq. (5) in the spaces parallel and perpendicular to $u^a$, respectively. Instead, Eqs. (9) and (10) are *off-shell* identities derived from the fact that the background spacetime is flat, and that the temperature is a scalar field. Finally, Eqs. (11) and (12) are just the definitions of $A$ and $S^a$ given in (2) and (6), respectively.

The authors studied three properties of the system (7)-(12): linear stability, causality and well-posedness. The fulfillment of the first two conditions is essential for the theory to provide a consistent and physically viable description of relativistic fluid dynamics [43, 80]. In order to ensure them, the authors imposed constraints for the transport coefficients, in particular requiring that $\chi = a_1\eta$ and $\lambda = a_2\eta$, where $\eta$ is analytic and $a_1$, $a_2$ are positive constants satisfying the inequalities [65]

$$a_1 > 4\,, \qquad a_2 \geq \frac{3a_1}{a_1-1}\,. \tag{13}$$

For proving local well-posedness, BDNK applied the Schauder method [59] (see also [81, 82]). Using the above causality conditions, BDNK equations are of CK type and the CK theorem can be applied to show existence of solutions. A key point afterwards is to derive energy estimates, for which the first-order reduction was used. By giving uniform Sobolev estimates for the difference between solutions, uniqueness was shown and, with a little more work, continuous dependence with initial data was also guaranteed.

The proof carried out by BDNK does not require propagating the constraints, but this should be a consequence of it. From a numerical perspective, the stability of the constraints along evolution is crucial for obtaining accurate evolutions, and this is why we believe that the study of their properties is relevant, constituting the core of the present work.

## III. CONSTRAINT PROPAGATION FOR THE CONFORMAL BDNK THEORY

We now study analytically the propagation of the algebraic and differential constraints of system (7)-(12), within the covariant framework described in the previous section. First, we derive a homogeneous system of differential equations ruling their evolution, and then we show that the latter is strongly-hyperbolic. Since the system will be shown to be homogeneous, the result presented here is enough to prove that, if all the constraints are satisfied at the initial time, then they will also be satisfied along the whole evolution.

### A. Constraint evolution system

By construction, the system (7)-(12) is equipped with a set of algebraic and differential conditions the evolution variables must satisfy along evolution, once they are properly prescribed at the initial time. For instance, the four-velocity $u^a$ is assumed to be of unit norm; that is, $u^a u_a + 1 = 0$. But since $u^a$ is a dynamical variable governed by Eq. (12), nothing guarantees, *a priori*, that such condition will be preserved along evolution. The same argument applies to the orthogonality between $u^a$ and $Q^a$, $S^a$ or $S^{ab}$.

In fact, the complete set of *algebraic constraints* is the following:

$$\begin{aligned}
C_1 &:= u^a u_a + 1 = 0\,; &(14)\\
C_2 &:= u^a Q_a = 0\,; &(15)\\
C_3 &:= u^a S_a = 0\,; &(16)\\
C_4^a &:= u^c {S_c}^a = 0\,; &(17)\\
C_5^a &:= u^c {S^a}_c = 0\,. &(18)
\end{aligned}$$

Notice that $C_4^a$ and $C_5^a$ represent different quantities, as the tensor $S_{ab}$ introduced in (6) is not necessarily symmetric. On top of the above listed algebraic constraints, there are also two *differential constraints* to be satisfied throughout the whole evolution, namely

$$\begin{aligned}
C_6^a &:= \frac{1}{\theta}\Pi^{ac}\nabla_c\theta + u^c\nabla_c u^a - \frac{Q^a}{\lambda} = 0\,, &(19)\\
C_7^{ab} &:= \Pi^{ac}\nabla_c u^b - S^{ab} = 0\,. &(20)
\end{aligned}$$

The first one corresponds to the definition of heat flux, given by Eq. (3), while the second one is defined in Eq. (6). Thus, what we need to prove is: if the quantities $\{C_1, C_2, C_3, C_4^a, C_5^a, C_6^a, C_7^{ab}\}$ are initially set to zero, then, as a consequence of Eqs. (7)-(12), they must be zero for all subsequent times. In order to see so, we first need to understand how the constraints (14)-(20) change in time; i.e., we need differential equations governing their evolution. Indeed, by taking derivatives along $u^a$, using the equations (7)-(12), and suitably re-expressing the lower-order terms, one can show that (14)-(20) satisfy the following evolution system:

$$\begin{aligned}
u^a\nabla_a C_1 + v_1 &= 0\,, &(21)\\
u^a\nabla_a C_2 - \eta\bar{\Pi}^a{}_b\nabla_a C_5^b + v_2 &= 0\,, &(22)\\
u^a\nabla_a C_3 - \frac{1}{\lambda}u^a\nabla_a C_2 + v_3 &= 0\,, &(23)\\
u^c\nabla_c C_4^a + v_4^a &= 0\,, &(24)\\
u^c\nabla_c C_5^a - \bar{\Pi}^{ab}\nabla_b C_3 + v_5^a &= 0\,, &(25)\\
u^c\nabla_c C_6^a + v_6^a &= 0\,, &(26)\\
u^c\nabla_c C_7^{ab} + v_7^{ab} &= 0\,, &(27)
\end{aligned}$$

where we have redefined the projector to the space orthogonal to $u^a$ in a more general way, namely

$$\bar{\Pi}^a{}_b := \delta^a{}_b + \frac{u^a u_b}{1 - C_1},$$

and such that $\bar{\Pi}^a{}_b = \Pi^a{}_b$ when $C_1 \equiv 0$; that is, when the unit-norm constraint holds. Similarly to the treatment of the full evolution system (7)-(12), the functions $v_i = v_1, \cdots, v_7$ depend algebraically on both constraints and dynamical fields, and their explicit form is given in Appendix A. Moreover, we notice at this point that the system of constraints is *homogeneous*, as each $v_i$ depends either linearly or quadratically on them. Thus, $C_1 = C_2 = C_3 = C_4^a = C_5^a = C_6^a = C_7^{ab} = 0$ *is a solution* of (21)-(27). Then, the last condition we need to show is that such solution is *unique*; i.e., if initially setting the whole set of constraints to zero would imply that they remain so for all times. This is shown in the following subsection.

### B. Strong hyperbolicity

In order to study the hyperbolicity of system (21)-(27), –an algebraic condition that guarantees existence and uniqueness of the solution given suitable initial conditions–, we find useful to express it in the form of a general first-order quasi-linear system; that is,

$$\mathcal{M}^{aA}{}_B \nabla_a C^B + V^A = 0\,, \tag{28}$$

which has dimension $31 \times 31$. Capital indices $A, B, \cdots$ denote internal indexing through the 31 constraints $C^A$,

$$C^A = \{C_1, C_2, C_3, C_4^a, C_5^a, C_6^a, C_7^{ab}\},$$

$V^A$ are the source functions

$$V^A = \{v_1, v_2, v_3, v_4^a, v_5^a, v_6^a, v_7^{ab}\},$$

and the *principal part* $\mathcal{M}^{aA}{}_B$ is explicitly given by

$$\mathcal{M}^{aA}{}_B = \begin{bmatrix} u^a & 0 & 0 & 0 & 0 & 0 & 0 \\ 0 & u^a & 0 & 0 & -\eta\bar{\Pi}^a{}_b & 0 & 0 \\ 0 & -\frac{u^a}{\lambda} & u^a & 0 & 0 & 0 & 0 \\ 0 & 0 & 0 & u^a & 0 & 0 & 0 \\ 0 & 0 & -\bar{\Pi}^{ba} & 0 & u^a & 0 & 0 \\ 0 & 0 & 0 & 0 & 0 & u^a & 0 \\ 0 & 0 & 0 & 0 & 0 & 0 & u^a \end{bmatrix}.$$

Then, the problem is reduced to showing that (28) is *strongly-hyperbolic* in a covariant framework. A natural first requirement for proving so is that, for any $t_a$ timelike covector, the condition

$$\det\left(t_a \mathcal{M}^{aA}{}_B\right) \neq 0$$

holds. In fact, this allows to choose a timelike direction and rewrite the system (28) in the more familiar ("3+1") way

$$\partial_t C^A + A^{iA}{}_B \partial_i C^B + \tilde{V}^A = 0\,, \tag{29}$$

where $A^{iA}{}_B := \left(t_a \mathcal{M}^{aA}{}_C\right)^{-1}\left(\mathcal{M}^{iC}{}_B\right)$ and $\tilde{V}^A := \left(t_a \mathcal{M}^{aA}{}_B\right)^{-1} V^B$. Then, strong-hyperbolicity would follow by showing that, for any spatial covector $w_i$, the symbol $w_i A^{iA}{}_B$ is *diagonalizable* with *real eigenvalues*. This is completely equivalent to prove that the *generalized eigenvalue problem*

$$\left(w_a - w\, t_a\right) \mathcal{M}^{aA}{}_B W^B = 0$$

admits a complete set of eigenvectors $\{W^B\}$ with purely real eigenvalues $w$, for all $t_a$ timelike and $w_a$ spacelike covectors.

Indeed, the characteristic polynomial

$$p(w) = \det\left[\left(w_a - w\, t_a\right) \mathcal{M}^{aA}{}_B\right]$$

depends on both $t_a$ and $w_a$. By direct computation, it is straightforward to see that $p(w) \equiv 0$ *if and only if*

$$\left[u^a \left(w_a - w\, t_a\right)\right]^{29} \Big\{\left[u^a \left(w_a - w\, t_a\right)\right]^2 - \frac{\eta}{\lambda}\bar{\Pi}^{ab}\left(w_a - w\, t_a\right)\left(w_b - w\, t_b\right)\Big\} = 0\,. \tag{30}$$

Solving (30) for $w$, we get *three* different solutions:

$$w_{29} = \frac{u^a w_a}{u^c t_c}\,;$$

$$w_{\pm} = \frac{E_{wt} \pm \sqrt{E_{wt}^2 - E_{ww}E_{tt}}}{E_{tt}}\,,$$

where we have denoted $E_{XY} := E^{ab} X_a Y_b$, with

$$E^{ab} = u^a u^b - \frac{\eta}{\lambda}\bar{\Pi}^{ab}\,.$$

Since $t_a$ is timelike and $w_a$ is spacelike, we get that $E_{wt}^2 > E_{ww}E_{tt}$ *if and only if* $\eta < \lambda$. This condition is in agreement with the general conditions (13) for the transport coefficients in the theorem 1.1 by BDNK [65]. Then, since $E_{tt} \neq 0$, the three eigenvalues are all real. Now, given that $w_+ \neq w_-$, the corresponding eigenspaces are one-dimensional. Finally, it is possible to show that the eigenspace corresponding to $w_{29}$ has dimension 29, as it corresponds to the null space of $\mathcal{N}^A{}_B = \left(w_a - w_{29}\, t_a\right)\mathcal{M}^{aA}{}_B$, which is a 29-dimensional vector space (see Appendix B for the details). Thus, the system (29) admits a complete set of eigenvectors, with purely real eigenvalues, the strong-hyperbolicity of (28) holds, and the constraints propagate correctly along evolution.

The above result is in agreement with the proof of local well-posedness given by BDNK using the first-order reduction (7)-(12), at it has direct applications. For instance, by using the constraint propagation it is possible to conclude that fluid configurations with uniform velocity but variable temperature are not physically sound, as conformal invariance would be broken. We give an argument of this fact in Appendix C.

From a numerical point of view, the inspection of the evolution of the constraints is necessary to guarantee accurate numerical simulations. However, the strong-hyperbolicity of the system of constraints does not imply that the equations are stable under small perturbations so, in principle, small discrepancies in the numerical approximations could lead to spurious modes that grow exponentially throughout the evolution. In what follows, we carry out a numerical exploration of this fact, evolving highly symmetric configurations, for which we study the propagation and stability of these constraints.

## IV. NUMERICAL SIMULATIONS

In this section we summarize the numerical set up implemented for our simulations, which are displayed in the next section. We restrict the BDNK system to plane-symmetric fluid configurations, getting a reduced number of evolution fields, and show the explicit equations, initial data and boundary conditions.

### A. Evolution variables

We start by deriving the explicit set of evolution equations from the covariant system (7)-(12). For doing so, we take a spatial hypersurface $\Sigma_o \simeq \mathbb{R}^3$, and cover it with global inertial coordinates $x^i = (x, y, z)$. We then pick a unit time-like vector field $t^a$, transverse to $\Sigma_o$, and extend the coordinates $x^i$ in a way that they are constant along the integral curves of $t^a$. Let $t : \mathbb{M} \to \mathbb{R}$ be a function which is zero on $\Sigma_o$ and satisfies $t^a \partial_a t = 1$. Then, we have that $t^a = (\partial/\partial t)^a$, and we choose $t$ to be the time coordinate. Under this construction, the fluid four-velocity can be expressed as [83]

$$u^a = \gamma(1, v^i)\,,$$

where $v^i$ is the 3-velocity of the fluid and

$$\gamma = \frac{1}{\sqrt{1 - v^i v_i}}$$

is the Lorentz factor. Furthermore, in order to simplify the scheme, we restrict our analysis to plane-symmetric (or *slab-symmetric*) configurations; i.e., we look for solutions that only depend on $t$ and $x$. Then, the fluid four-velocity is $u^a = \gamma\,(1, v, 0, 0)$, being $v(t, x)$ the corresponding velocity field.

For the numerical simulations, we need to rewrite the original covariant reduction (7)-(12) as a quasi-linear PDE system of the form

$$\partial_t \phi^\alpha + \mathcal{A}^\alpha{}_\beta\, \partial_x \phi^\beta + \mathcal{R}^\alpha = 0, \tag{31}$$

where $\phi^\alpha = \{A, Q^a, S^a, S_a{}^b, \theta, u^a\}$ is the set of dynamical fields, $\mathcal{A}^\alpha{}_\beta$ its principal part, and $\mathcal{R}^\alpha$ some linear combination of the source terms $r_i$ of Eqs. (7)-(12). For the fluid four-velocity, we choose to evolve the component $u^1 \equiv \gamma v$, from which the Lorentz factor can be expressed as

$$\gamma = \sqrt{1 + (u^1)^2}\,.$$

For the heat flux evolution, we use the planar symmetry and choose $Q^1$ as a dynamical variable. The component $Q^0$ can be directly retrieved from the orthogonality condition $Q^a u_a = 0$, and rewritten in terms of $u^1$ and $Q^1$ getting

$$Q^0 = \frac{Q^1 u^1}{\sqrt{1 + (u^1)^2}}.$$

An analogous procedure can be followed for the variables $S^a$ and $S^a{}_b$. In particular, due to the symmetries of the problem, it is possible to express the latter only in terms of one free field, which we choose to be $S^{11}$. Indeed, the conditions $S^{ab} u_b = 0$ and $S^{ab} u_a = 0$ imply that

$$S^{ab} = S^{11} \begin{pmatrix} \frac{(u^1)^2}{1+(u^1)^2} & \frac{u^1}{\sqrt{1+(u^1)^2}} & 0 & 0 \\ \frac{u^1}{\sqrt{1+(u^1)^2}} & 1 & 0 & 0 \\ 0 & 0 & 0 & 0 \\ 0 & 0 & 0 & 0 \end{pmatrix}.$$

We recall that the above symmetry found for $S^{ab}$ is a pure consequence of the planar symmetry assumed for this construction; i.e., for a general four-velocity $u^a$, the tensor $S^{ab}$ is *not* necessarily symmetric.

In summary, the set of evolution variables for our simulations becomes

$$\Phi = \{A, Q^1, S^1, S^{11}, \theta, u^1\}\,.$$

### B. Dynamical equations

In order to get with a system of evolution equations for the reduced set of variables $\Phi$, we need to find $\mathbb{A}(\Phi)$ and $\mathbf{R}(\Phi)$ such that the system (31) reduces to

$$\partial_t \Phi + \mathbb{A}(\Phi)\, \partial_x \Phi + \mathbf{R}(\Phi) = 0\,.$$

To do so, we split the covariant equations and use the planar symmetry. First, we notice that the equations for $\theta$ and $u^1$ are "decoupled" from the rest of the system, as they *only* contain spatial derivatives with respect to themselves. Such equations are obtained from Eqs. (11) and (12), and read

$$\partial_t \begin{pmatrix} u^1 \\ \theta \end{pmatrix} = -\frac{u^1}{\gamma} \begin{pmatrix} 1 & 0 \\ 0 & 1 \end{pmatrix} \partial_x \begin{pmatrix} u^1 \\ \theta \end{pmatrix} + \begin{pmatrix} R_u \\ R_\theta \end{pmatrix}, \tag{32}$$

where

$$\begin{aligned} R_u &= -\frac{S^1}{\gamma}\,, \\ R_\theta &= \frac{\theta}{3\gamma^3} \left[ S^{11} + \frac{A\left(1 - (u^1)^2\right)}{\chi} \right], \end{aligned}$$

and we set $\chi = \chi_0 \theta^3$, with $\chi_0 > 0$ a constant parameter. The explicit form of the remaining four evolution equations for the fields $\{A, Q^1, S^1, S^{11}\}$ were obtained by solving a linear system, following exactly the same procedure illustrated for $\{u^1, \theta\}$. The final form of the system is detailed in Appendix D. All the calculations were implemented in `Mathematica` [84], and here we omit the explicit solutions as they are quite intricate.

### C. Differential constraints

During the evolution of the dynamical fields, we monitor the value of the differential constraints (19) and (20), which were introduced in order to get the first-order reduction. Using the planar symmetry and the coordinate choice introduced before, we can express them as a function of our evolution variables, namely

$$
\begin{aligned}
Q^1 &= \lambda \left[ u^1 \partial_x u^1 + \gamma \partial_t u^1 + \frac{\gamma}{\theta} \left( \gamma \partial_x \theta + u^1 \partial_t \theta \right) \right], \\
S^{11} &= \gamma u^1 \partial_t u^1 + \gamma^2 \partial_x u^1 \, .
\end{aligned}
$$

Using Eqs. (32) for $u^1$ and $\theta$, we can replace the time-derivatives, finally getting

$$
\begin{aligned}
Q^1 &= \lambda \left( \frac{\partial_x \theta}{\theta} + S^1 + \frac{A u^1}{3\chi} - \frac{S^{11} u^1}{3\gamma^2} \right) && (33) \\
S^{11} &= S^1 u^1 + \partial_x u^1 \, , && (34)
\end{aligned}
$$

where $\lambda = \lambda_0 \theta^3$, with $\lambda_0 > 0$ a constant parameter. Appropriate initial conditions are then prescribed for the fluid fields such that the constraints (33) and (34) are satisfied at $t = 0$. As a consequence of the dynamical evolution, it is expected that the they remain constant for $t > 0$.

### D. Initial data, boundary conditions and implementation details

For the simulations, we consider initial gaussian profiles of the form

$$
f(x) = f_0 + f_1 \exp\left[ -\frac{(x - x_c)^2}{\sigma^2} \right],
$$

with center at $x_c = 0$ and width $\sigma = 2$. In order to explore the propagation of the constraints at different initial configurations, we considered four different data (**ID1**, **ID2**, **ID3** and **ID4**), varying the coefficient $f_1$ for $A$ and $S^1$ while keeping the rest of them unaltered. The coefficients $f_0$ and $f_1$ depend on each field, and are listed in Table I. Once the initial data for $\{u^1, \theta, A, S^1\}$ has been set, we prescribe the corresponding data for $\{Q^1, S^{11}\}$ by means of the equations (33) and (34), ensuring that the constraints are initially satisfied.

For the computations, we take a uniform grid from $x_0 = -100$ to $x_f = 100$, and $N = 20001$ grid points, for which the grid size is $\Delta x = 0.01$. For the time step, we took $\Delta t = 0.0025$. We discretized the spatial derivatives using centered finite differences with second-order accuracy and integrated forward in time by means of a standard fourth-order Runge-Kutta method. In order to simplify the scheme, we imposed periodic boundary conditions, for which we set a large enough numerical domain so as to avoid spurious boundary reflections during the evolution. To achieve stable simulations, we also included Kreiss-Oliger (K-O) artificial dissipation [85, 86], which removes high-frequency modes (i.e., modes with a wavelength smaller than the grid size $\Delta x$) that can not be accurately resolved in the grid. In particular, we considered the 6th order K-O operator given by

$$
\begin{aligned}
(\mathcal{D} f)_i &= \frac{\sigma_{\text{diss}}}{64 \Delta x} \left( f_{i-3} - 6 f_{i-2} + 15 f_{i-1} - 20 f_i \right. \\
&\quad \left. + 15 f_{i+1} - 6 f_{i+2} + f_{i+3} \right) , && (35)
\end{aligned}
$$

and took $\sigma_{\text{diss}} = 0.25$. The form of this operator has been shown to be very suitable for 4th-order integration schemes, as the one here implemented (see [87] for details). Finally, for the validation of our code, we performed convergence tests in both time and space. The results are displayed in the following section.

| | | **ID1** | **ID2** | **ID3** | **ID4** |
|---|---|---|---|---|---|
| Field | $f_0$ | $f_1$ | $f_1$ | $f_1$ | $f_1$ |
| $v$ | 0.3 | 0.1 | 0.1 | 0.1 | 0.1 |
| $\theta$ | 300 | 10 | 10 | 10 | 10 |
| $A$ | 0 | 0 | 0 | 0.5 | 0.5 |
| $S^1$ | 0 | 0 | 0.5 | 0 | 0.5 |

TABLE I. *Fluid initialization.* Gaussian parameters for the initial data. The component $u^1$ is computed from the 3-velocity $v$ as $u^1 = v/\sqrt{1 - v^2}$.

## V. RESULTS

### A. Dynamical evolution

Figure 1 shows the profiles of the six dynamical fields at three different evolution times: $t = 26$ (first column), $t = 48$ (second column), and $t = 80$ (third column), having initialized the code with **ID3**. We observe that all profiles display a smooth evolution, so up to $t_f = 80$ the fields do not develop discontinuities or shock waves. This was also the case for all the other initial configurations considered in this work. This is important when evaluating the propagation of the constraints as, since we are considering an explicit numerical scheme that is not suitable for capturing shocks, any discontinuity appearing in the evolution would produce small high-frequency oscillations, which would grow spuriously over time, thus affecting the convergence of the method. Then, any unexpected behavior in the propagation of the constraints during the evolution could be due to the lack of convergence of the implemented scheme. However, for our simulations, we chose initial data such that, up to the final time $t_f$ here considered, the evolution is perfectly smooth and, as we will report below, the convergence in both time and space is the expected one. Thus, all the necessary conditions for a proper constraint evolution are met.

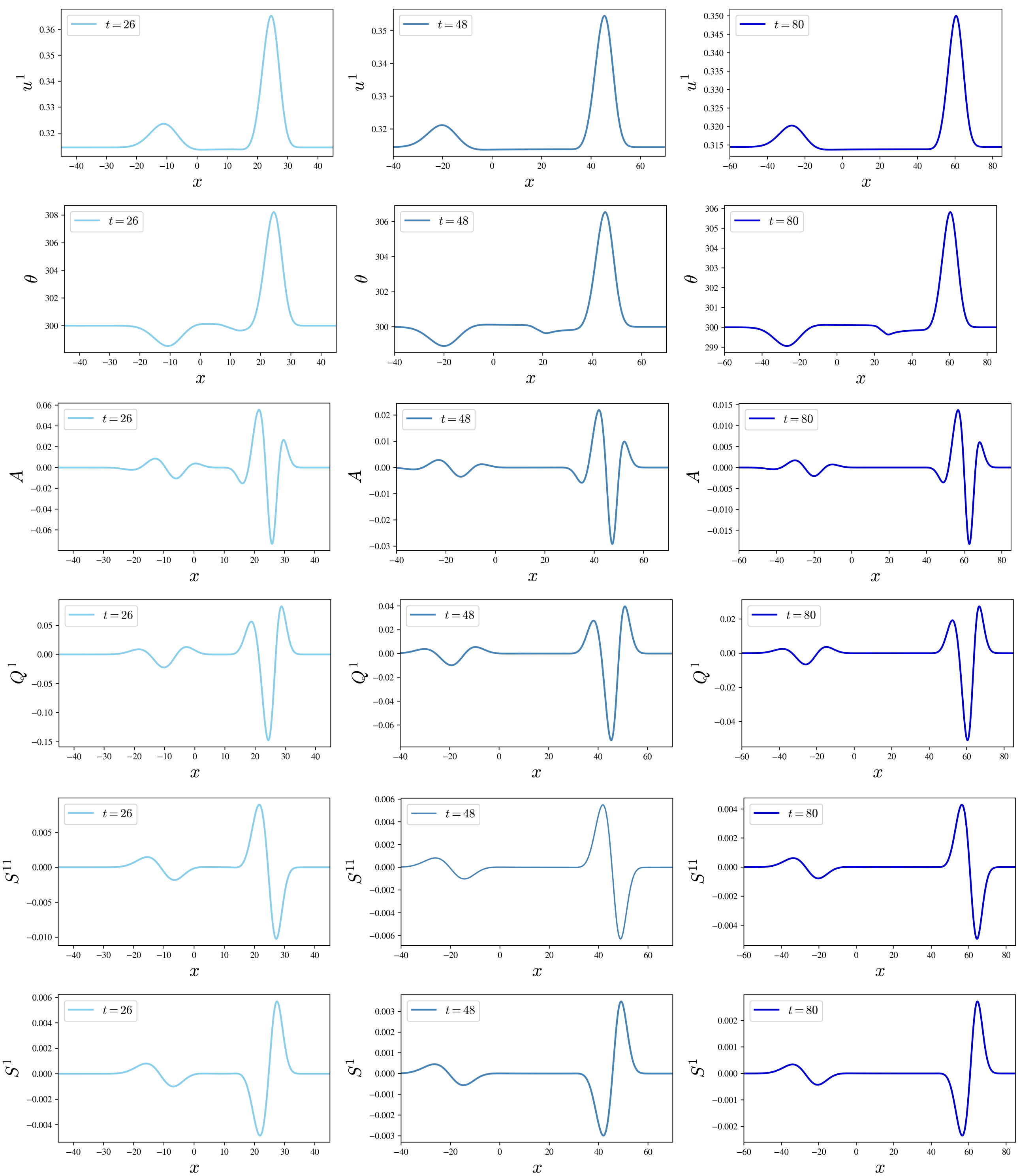

FIG. 1. *Fluid dynamics.* Snapshots of the evolution of the fluid fields at three different times: $t = 26$ (first column), $t = 48$ (second column), and $t = 80$ (third column), starting from the initial data **ID3** specified in Table I. All the fields display smooth evolution, without shock formation. The simulations were done with an explicit 4th-order Runge-Kutta scheme, and the method of lines for the spatial grid discretization. For the simulations, we took $N = 20001$ grid points, with $\Delta x = 0.01$ and $\Delta t = 0.0025$.

### B. Propagation of the constraints

As anticipated in the previous section, during the fluid evolution we controlled the propagation of the constraints. For doing so, we define the following quantities:

$$C_{S^{11}} \equiv S^{11} - S^{11}(\Phi, \partial_x \Phi)\,; \tag{36}$$
$$C_{Q^1} \equiv Q^1 - Q^1(\Phi, \partial_x \Phi)\,, \tag{37}$$

where the functions $S^{11}(\Phi, \partial_x \Phi)$ and $Q^1(\Phi, \partial_x \Phi)$ are given explicitly by the right-hand sides of equations (33) and (34), respectively. Then, by construction, the initial value of both constraints is identically zero. During the subsequent evolution we expect both quantities to remain close to zero; i.e., to display no appreciable growth.

Since $C_{S^{11}}$ and $C_{Q^1}$ are not only function of time but also of the coordinate $x$, we study the evolution of the $L^2$ norm, which in their discrete versions read [88]

$$\|C_J\|_{L^2}(t) = \frac{1}{N}\sqrt{\sum_{i=1}^{N} C_J(t, x_i)^2}\,.$$

Figure 2 shows the results obtained for the evolution of $C_{S^{11}}$ (left panel) and $C_{Q^1}$ (right panel), for the four different initial data here considered. As it can be seen from the left panel, there is a transient time in which the $L^2$ norm of $C_{S^{11}}$ displays an oscillatory behavior, but then it settles down, decaying to a value close to zero. For the case of $C_{Q^1}$ shown in the right panel, even though the value of its $L^2$ norm is a few orders of magnitude higher than the one for $S^{11}$, it exhibits a constant evolution since the beginning of the simulation.

The above results allow for analyzing the stability of the constraints under numerical uncertainties. As it is expected, the derivatives involved for computing the initial data are evaluated using centered finite differences, whose error is of order $\mathcal{O}(\Delta x^2)$, implying a small departure from the theoretical values for the initial configuration. This means that, even though the evolution of the constraints admits a unique solution, and such a solution is *exactly zero* choosing *zero* initial data, nothing prevents that numerical errors at the initial data could lead to exponential growth of them; that is, small initial perturbations around zero could display unstable modes. This can be easily derived from the structure of the evolution equations governing the dynamics of the constraints, which have the general form $u^a \partial_a C^A \propto D^A{}_B C^B$. Then, depending on the sign of $D^A{}_B$, the equations may have exponentially growing spurious modes. In our particular case, as we are only analyzing the evolution of the *differential* constraints (those algebraic are by construction guaranteed from the variable reduction for the numerical evolution), we should look for the Eqs. (26) and (27), leaving only the terms proportional to $C_6$ and $C_7$ (which correspond to $C_{Q^1}$ and $C_{S^{11}}$ respectively). By a direct inspection on them, we see that they do not have any definite sign, meaning that it could change in time but also in space. From the simulations here displayed we see that, even though the initial data may admit numerical discrepancies, their subsequent evolution does not exhibit any spurious mode, making them to settle down towards values very close to zero. We confirm that the same behavior occurs when varying $f_0$ and $\sigma$, here reporting an illustrative example.

Thus, the first-order reduction for the BDNK formulation suggested in Eqs. (7)-(12) seems to be adequate not only for showing local well-posedness of the conformal theory, but also for reaching stable numerical simulations.

### C. Robustness of the results

#### 1. *Convergence tests*

In order to assess the validity of our numerical scheme, we performed several convergence tests, verifying the accuracy of the methods here used. Indeed, we displayed convergence tests both in time and space domains. As it was pointed out in the previous sections, for the time integration we implemented a fourth-order Runge-Kutta scheme, adding artificial viscosity for avoiding spurious oscillations near eventual shock fronts. Instead, for the spatial discretization, we implemented the method of lines by means of centered finite differences which were chosen to be second-order accurate. On top of this, our finite differences satisfy the property of summation by parts [86], allowing semi-discrete energy estimates for the corresponding initial-boundary value problem, thus guaranteeing numerical stability.

If the numerical scheme is of order $p$, we can assess its correct operation by performing three runs at different resolutions; say $r$, $r/2$ and $r/4$. Then, define the (logarithmic) convergence factor [88]

$$Q(t) \equiv \log_2\left(\frac{\|\Phi_r(t,x) - \Phi_{r/2}(t,x)\|}{\|\Phi_{r/2}(t,x) - \Phi_{r/4}(t,x)\|}\right),$$

where $\Phi_r$, $\Phi_{r/2}$ and $\Phi_{r/4}$ represent the evolutions obtained at resolutions $r$ (low), $r/2$ (medium) and $r/4$ (high), respectively, while $\|\cdot\|$ is some adequate norm taken over the whole spatial domain. It can be shown that if the method is order $p$, then it should hold $Q(t) \sim p$. This standard way to test the accuracy of a numerical scheme is useful for (at least) three important reasons: (i) the convergence factor $Q(t)$ is independent on the initial chosen resolution $r$; (ii) this analysis does not need an "exact" solution for a correct assessment (unlike other convergence tests, like independent residual evaluators);

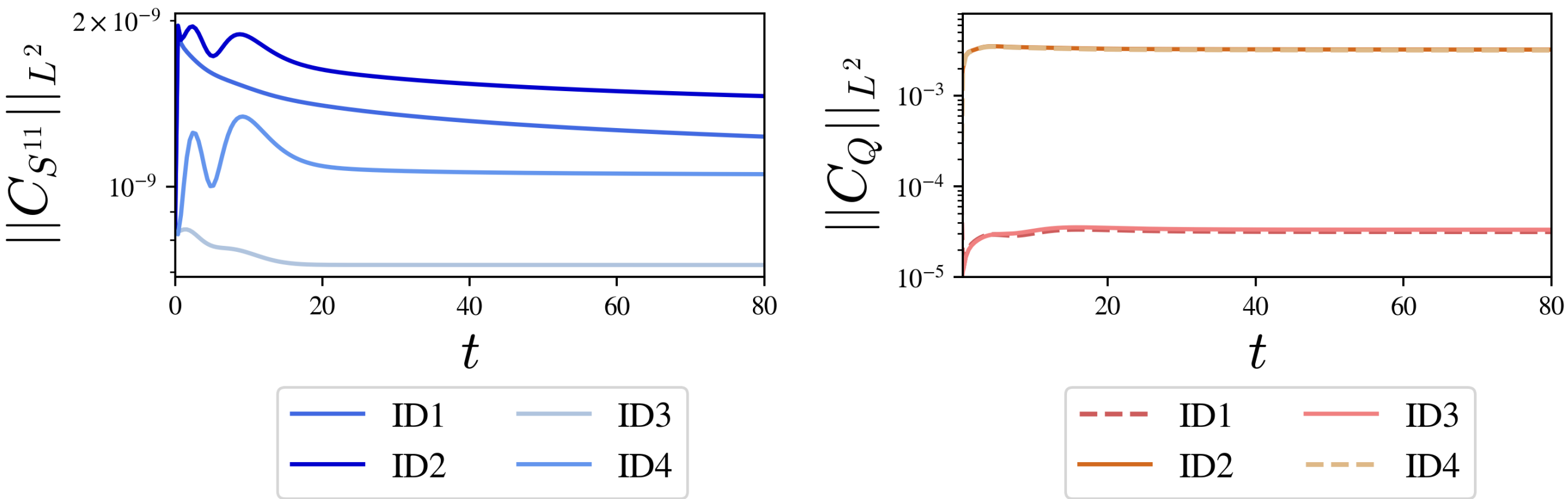

FIG. 2. *Constraint propagation.* Time evolution of the $L^2$ norms of the constraints $C_{S^{11}}$ (left panel) and $C_{Q^1}$ (right panel) given in Eqs. (33) and (34) for all the initial configurations considered. In all the cases, the constraints reach a stationary value after a short transient time. For $C_{S^{11}}$, a smooth decay to values close to zero is observed, while $C_{Q^1}$ displays a value a few orders of magnitude higher, exhibiting anyway a constant evolution since the beginning of the simulation.

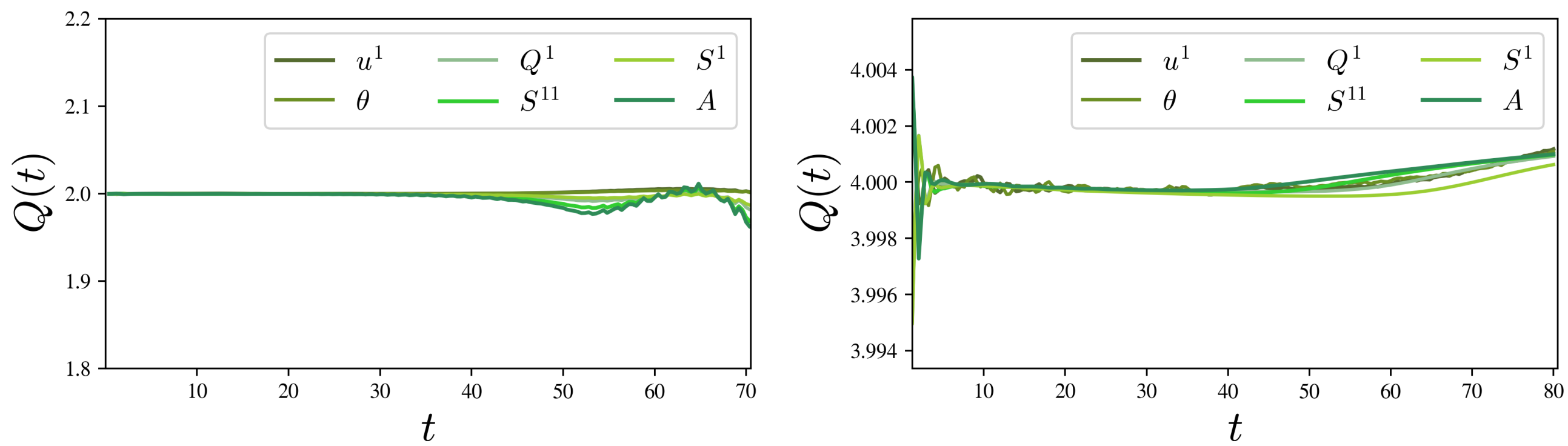

FIG. 3. *Robustness of the simulations.* Precision factors for convergence in space (left panel) and time (right panel), as a function of time, for each one of the dynamical fields. For the convergence in time, we took $N = 10001$ grid points and $\Delta t = 0.008$ as the lowest time step. For the convergence in space, we took $N_1 = 8001$ grid points for the lowest resolution, while fixed the time step to $\Delta t = 0.001$. Both tests yielded the expected results.

and (iii) it may be performed *locally* in time; i.e., assuring the expected result at a very first initial time is enough to guarantee the correct operation of the scheme. Usually, it is performed along the whole time evolution, for assuring that the convergence is not being lost for physical reasons. Nonetheless, an adequate norm needs to be chosen. For our purposes, as we are dealing with smooth evolutions, the $L^2$ norm is suitable for this test. For shock propagation, instead, in addition to considering shock-capturing schemes, a $L^1$ norm would be more adequate, or even more sophisticated ones, like the "Lip′" family of norms (for details, see Refs. [89, 90]).

For our tests, we assessed the convergence and stability both in time and space. For the time convergence, we fixed the number of grid points to $N = 10001$, performed three runs with time steps $\Delta t = 0.008$ (low resolution), $\Delta t/2$ (medium resolution) and $\Delta t/4$ (high resolution), and compared the obtained accuracy order with the expected nominal value, which in this case should be $\sim 4$. We did this analysis for all the dynamical fields, obtaining similar results, as it can be seen in the right panel of Figure 3. The analogous result for convergence in space is shown in the left panel. In this second test, we instead fixed the time step to $\Delta t = 0.001$, considered three different spatial resolutions: $N_1 = 8001$ (low), $N_2 = 16001$ (mid) and $N_3 = 32001$ (high), and computed the same convergence ratio. In this case, the result yields the accuracy order of the chosen finite difference operator, and is reported in the left panel of the figure, for all the dynamical fields. This analysis allows us to trust in the robustness of our results, as well as to draw solid conclusions on the propagation of the constraints.

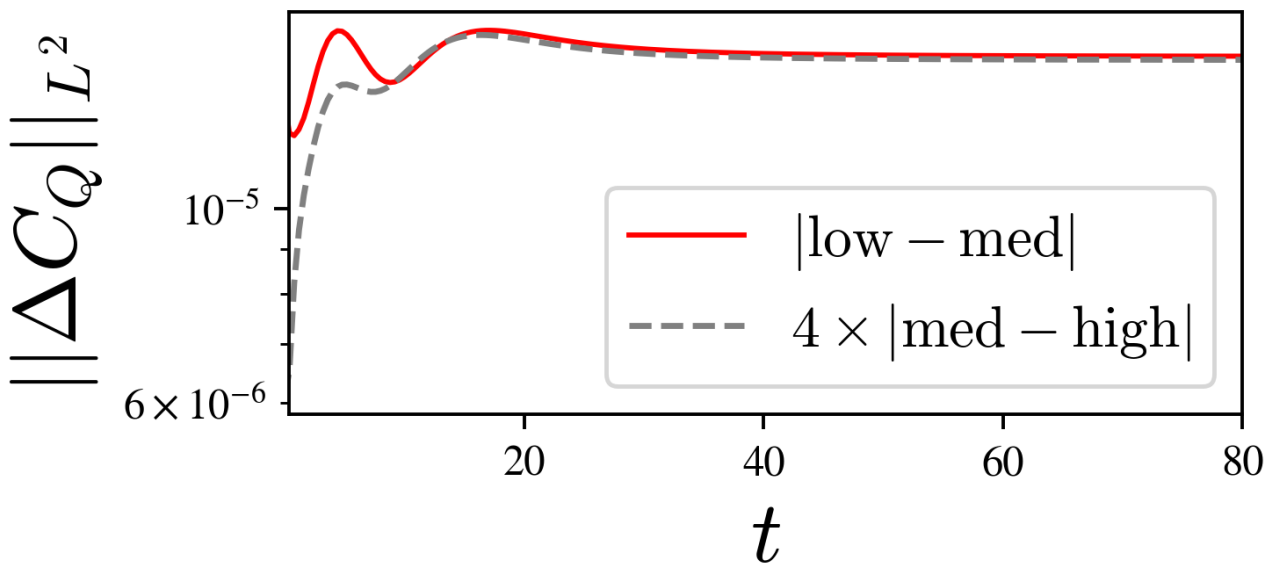

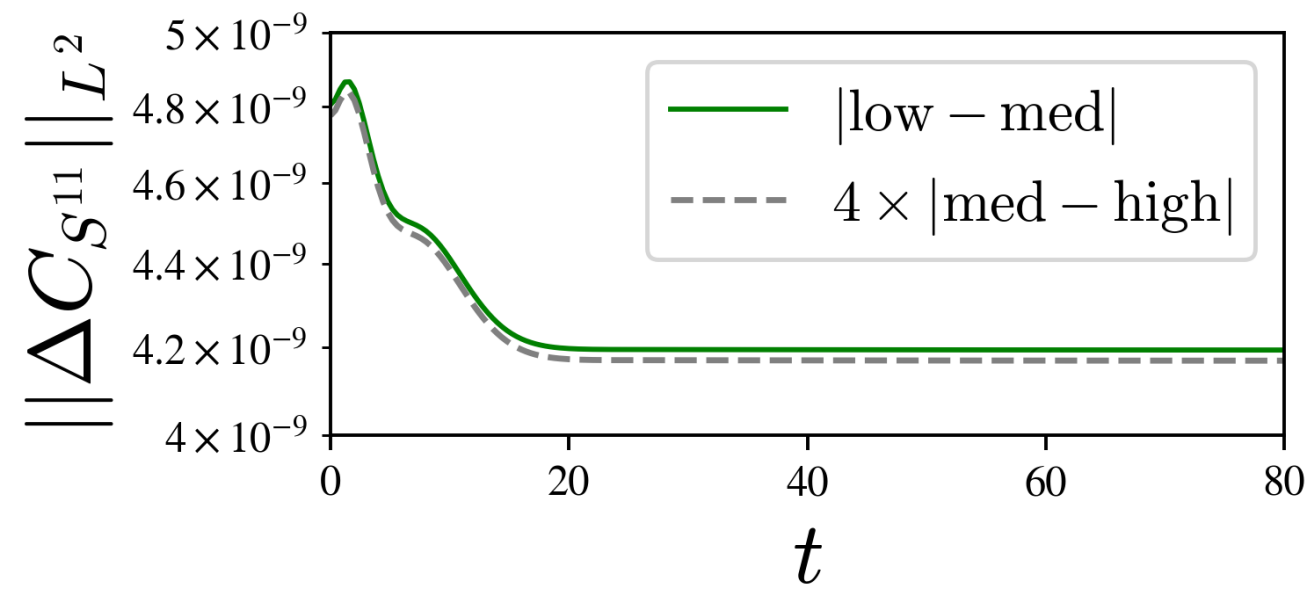

FIG. 4. *Convergence of the constraints.* Absolute value of the difference between the low and medium resolutions approximation (continuous curves) and between the medium and high ones (dashed grey curves) for $C_{Q^1}$ (left) and $C_{S^{11}}$ (right). The simulatios were initialized with ID3, and numerical integration were done with lowest spatial resolution $\Delta x = 0.02$, keeping $\Delta t = 0.001$. The dashed curve is shifted by a factor 4, confirming second-order accuracy for the convergence, as expected.

#### 2. *Convergence of the constraints*

As a final consistency check of our scheme, we performed a convergence test for the differential constraints, in a similar way that is was done for the evolution variables, with the scope to quantify how better the constraints propagate when increasing the numerical resolution (and thus, decreasing the numerical truncation error). Since we are using second order accurate finite difference operators, given a spatial resolution $\Delta x$, the numerical approximation of the constraints should scale as

$$C_{\text{num}} \sim 0 + \alpha(\Delta x)^2,$$

with $\alpha$ depending on the numerical scheme. Then, by computing the numerical approximations $C^{(1)}_{\text{num}}$, $C^{(2)}_{\text{num}}$ and $C^{(3)}_{\text{num}}$ with respective resolutions $\Delta x_1$, $\Delta x_2$ and $\Delta x_3$, we should get

$$\frac{\|C^{(2)}_{\text{num}} - C^{(1)}_{\text{num}}\|}{\|C^{(3)}_{\text{num}} - C^{(2)}_{\text{num}}\|} \sim \left| \frac{1 - \left(\frac{\Delta x_1}{\Delta x_2}\right)^2}{\left(\frac{\Delta x_3}{\Delta x_2}\right)^2 - 1} \right| .$$

In particular, for resolutions $\Delta x$, $\Delta x/2$, and $\Delta x/4$, the above quotient should be

$$\frac{\|C^{(\text{med})}_{\text{num}} - C^{(\text{low})}_{\text{num}}\|}{\|C^{(\text{high})}_{\text{num}} - C^{(\text{med})}_{\text{num}}\|} \sim 4 .$$

We verified that this condition actually holds for our scheme, by performing numerical integrations with lowest resolution $\Delta x = 0.02$, keeping the time step to be always $\Delta t = 0.001$. The result is shown in Figure 4, in which we compare the $L^2$ norm of the difference between the low and med approximations, with the one between the med and high ones, showing the expected approximated scaling behavior.

## VI. FINAL REMARKS

In this work we have investigated the set of constraints, both algebraic and differential, that arise from the first-order reduction proposed by BDNK for the study of the initial-value problem of first-order relativistic viscous fluids with conformal invariance. This reduction was introduced by BDNK to prove, using Schauder's method and certain energy estimates, that this theory is locally well-posed in Sobolev spaces. Here we complement their study by means of the numerical counterpart, with the aim of applying it to the evolution of fluid systems of astrophysical interest. From a numerical point of view, the preservation of the constraints is an essential condition in order to guarantee stable and accurate simulations. Thus, it is not only necessary to demonstrate that the former are consistent with the evolution equations, but it is also crucial to prove that there are no spurious modes that develop exponential growth during the numerical evolution of the system.

Firstly, we studied the initial-value problem of the complete set of constraints, derived from the BNDK reduction. Using the evolution equations of the conformal fluid, we found that the constraints satisfy a homogeneous system of first-order quasi-linear equations. As a consequence of the homogeneity, we concluded that there exists a solution which is consistent with the fluid equations. Then, by means of standard hyperbolicity methods, we saw that the principal part is diagonalizable with real eigenvalues, implying that the system of constraints is strongly-hyperbolic, therefore also admitting a locally well-posed formulation. In conclusion, and in line with the original proof of local well-posedness provided by BDNK in [65], the constraints are preserved during the evolution: if they hold at the initial time, then they are also satisfied afterwards.

The above result, however, does not imply that such solution for the constraints remains bounded under small perturbations coming, for instance, from numerical truncation errors. This second issue motivated an deep exploratory numerical study of how these constraint con-ditions are preserved during evolution. To this end, we performed numerical simulations by restricting the entire system imposing slab symmetry, and using the algebraic constraints to reduce the number of evolution variables. The simulations were initialized by field configurations satisfying the differential constraints, and during the evolution we monitored their $L^2$-norm. We report that, in all the study cases considered in this work, the constraints reach a stationary value, after a short transient time, which remains stable throughout the simulation. Although we observed that numerical errors lead $C_{Q^1}$ settling to a value a few orders of magnitude greater than the one reached by $C_{S^{11}}$, both remained constant and stable during the whole evolution. In order to validate our scheme, several convergence testes were performed, in space a time, not only for the dynamical variables but also for the scaling of the constraints when increasing the resolution. In all of them we obtained the expected behavior, supporting the robustness of the scheme.

Finally, and in light of the results presented in this work, we expect to implement a similar reduction of order and constraint propagation for solving the fluid equations in a fully flux-conservative formulation, in order to deal with shock formation as a consequence of the nonlinear character of the equations. This would imply, in principle, finding conserved variables and rewriting the BDNK system in conservative form, but keeping the spirit of such reduction. In particular, we expect to adapt the same ideas to a curved background, for instance by means of coupling the fluid theory to Einstein's field equations. This could be challenging, as the reduced fluid system considers commutation of covariant derivatives, which would lead to terms proportional to the Riemann tensor, being of leading-order when coupled to the field equations.

## ACKNOWLEDGEMENTS

We are grateful to Áron Kovács, Luis Lehner and Oscar Reula for enlightening discussions throughout the initial stages of this project. We would also like to thank Fernando Abalos, Miguel Bezares and Harry Shum for their useful comments on a first version of the manuscript. Finally, we are deeply grateful to the anonymous referees for their valuable suggestions on the proof of the constraint propagation as well as for pointing out some conformal invariance considerations. MR acknowledges support from PRIN 2022 grant "GUVIRP - Gravity tests in the UltraViolet and InfraRed with Pulsar timing", the EU Horizon 2020 Research and Innovation Programme under the Marie Sklodowska-Curie Grant Agreement No. 101007855 and the MUR PRIN Grant No. 2022-Z9X4XS funded by the European Union (Next Generation EU).

## Appendix A: Lower-order terms of the evolution systems

In this appendix we show the explicit expressions for both the functions $r_i$ and $v_i$ corresponding to the lower-order terms (LOTs) of the full evolution system (7)-(12) and the system of constraints, respectively.

### 1. LOTs of the full evolution system

The functions $r_i = r_1, \cdots, r_6$ are to be written only as algebraic functions of the evolution variables $\{A, Q^a, S^a, S_a{}^b, \theta, u^a\}$, and were computed to be

$$r_1 = \frac{4}{3}AS + Q^a S_a - \frac{1}{2}\eta\sigma^2 + \frac{4\varepsilon_o}{3\chi}A\,\theta^4\,; \tag{A1}$$

$$\begin{aligned} r_2^a = \; & 4AS^a - \eta\left[3S^c S_c{}^a - 2SS^a - 3S_b{}^c S_c{}^b u^a\right] \\ & + 9\eta\sigma^{ab}S_b - 3Q^b S_b u^a + 3SQ^a \\ & + 3Q^b S_b{}^a + \frac{4\varepsilon_o}{\lambda}\theta^4 Q^a - \frac{9\eta}{\lambda}\sigma^{ab}Q_b\,; \end{aligned} \tag{A2}$$

$$r_3^a = \left(S - \frac{A}{\chi}\right)S^a + 3\left(u^a S^b - S^{ab}\right)\left(S_b - \frac{Q_b}{\lambda}\right)\,; \tag{A3}$$

$$r_4^{ab} = -S^a S^b - S^c S_c{}^a u^b + S_c{}^a S^{bc}\,; \tag{A4}$$

$$r_5 = -\frac{A}{3\chi}\,; \tag{A5}$$

$$r_6^a = -S^a\,; \tag{A6}$$

where $S = g_{ab}S^{ab}$, $\sigma^2 = \sigma^{ab}\sigma_{ab}$ and the shear is

$$\sigma_{ab} = S_{ab} + S_{ba} - \frac{2}{3}S\,\Pi_{ab}. \tag{A7}$$

Also, following the definition of the tensor $B_a{}^{bcd}$ given by the authors right after their Eq. (2.2f) in [65], we have the identity

$$\begin{aligned} B_a{}^{bcd}\nabla_d S_c{}^a \; = \; & -3\eta\nabla_a\sigma^{ab} \\ & + \; \eta\left[3S^a S_a{}^b - 2SS^b + \left(3S_a{}^c S_c{}^a - 2S^2\right)u^b\right]. \end{aligned}$$

### 2. LOTs of the constraint evolution system

For the case of the functions $v_i = v_1, \cdots, v_7$, they should be algebraic functions of the evolution variables $\{A, Q^a, S^a, S_a{}^b, \theta, u^a\}$ but also of the constraints $C_i$, which were introduced in Eqs. (14)-(20). The explicit

form of them is

$$
\begin{aligned}
v_1 &= -2C_3\,; && \text{(A8)}\\
v_2 &= \left(\eta S_{ab}S^{ab} - Q^a S_a\right) C_1 + \left(S + \frac{4\epsilon}{3\lambda}\right) C_2 \\
&\quad + \left(A + \frac{2}{3}\eta S\right) C_3 - 3\eta\left(\frac{Q_a}{\lambda} - S_a\right) C_4^a \\
&\quad + \left[2\eta S_a + \left(1 - \frac{3\eta}{\lambda}\right) Q_a\right] C_5^a + \eta S_{ab} C_7^{ab}\,; && \text{(A9)}\\
v_3 &= S^a\left(\frac{Q_a}{\lambda} - S_a\right) C_1 \\
&\quad + \left[\frac{3}{\lambda^2}\left(\frac{A}{\chi} - S\right) - \frac{A}{\lambda\chi^2}\right] C_2 \\
&\quad + \left[\frac{A}{\chi^2} + \frac{1}{3}\left(\frac{A}{\chi} - S\right)\right] C_3 \\
&\quad - \left(\frac{Q_a}{\lambda} - S_a\right) C_4^a\,; && \text{(A10)}\\
v_4^a &= -S^c S_c{}^a C_1 - S^a C_3 + S_c{}^a C_4^c\,; && \text{(A11)}\\
v_5^a &= -S^a C_3 + \left(S_c{}^a - u^a S_c\right) C_5^c + S_b C_7^{ab}\,; && \text{(A12)}\\
v_6^a &= \frac{u^a}{3}\left(\frac{A}{\chi} - S\right) C_3 \\
&\quad + \left(\frac{A}{\chi^2}\delta^a{}_b + S^a{}_b - u^a S_b\right) C_6^b \\
&\quad + \left(C_b^6 + \frac{Q_b}{\lambda} - S_b\right) C_7^{ab}\,; && \text{(A13)}\\
v_7^{ab} &= u^a S^b C_3 - S^b C_5^a + \left(S^a{}_c - u^a S_c\right) C_7^{cb} \\
&\quad + \left[S_c{}^b - u_c S^b + (C_7)_c{}^b\right] C_7^{ac}\,, && \text{(A14)}
\end{aligned}
$$

where we have followed the same conventions and notations used in the previous subsection and in Section II. We notice that each LOT depends either linearly or quadratically on the constraints $C_i$, and thus the system (14)-(20) is homogeneous.

## Appendix B: The eigenspace of $w_{29}$

In this appendix we explicitly compute the eigenspace associated with the eigenvalue $w_{29}$, computed in Section III. By definition, such space corresponds to the null space of the matrix $\mathcal{N}^A{}_B = (w_a - w_{29}\, t_a)\,\mathcal{M}^{aA}{}_B$. Noting that $u^a\,(w_a - w_{29}\, t_a) = 0$ and defining $n_b = -\bar{\Pi}^a{}_b\,(w_a - w_{29}\, t_a)$, we get

$$
\mathcal{N}^A{}_B = \left[\begin{array}{c|c}
\mathbf{0}^{7\times 7} & \begin{matrix} 0 & 0 & 0 \\ \eta n_b & 0 & 0 \\ 0 & 0 & 0 \\ 0 & 0 & 0 \end{matrix} \\
\hline
\begin{matrix} 0 & 0 & n^b \\ 0 & 0 & 0 \\ 0 & 0 & 0 \end{matrix} & \mathbf{0}^{24\times 24}
\end{array}\right].
$$

The corresponding null space can be parametrized in terms of any set of three vectors $\{n_\perp^1, n_\perp^2, n_\perp^3\}$ orthogonal to $n^b$, and is explicitly given by

$$
\begin{aligned}
\text{null}\left(\mathcal{N}^A{}_B\right) = \text{span}\Bigg\{ &\begin{pmatrix} 1 \\ \mathbf{0}^{30\times 1} \end{pmatrix}, \begin{pmatrix} 0 \\ 1 \\ \mathbf{0}^{29\times 1} \end{pmatrix}, \cdots, \begin{pmatrix} \mathbf{0}^{6\times 1} \\ 1 \\ \mathbf{0}^{25\times 1} \end{pmatrix}, \\
&\begin{pmatrix} \mathbf{0}^{7\times 1} \\ n_\perp^1 \\ \mathbf{0}^{20\times 1} \end{pmatrix}, \begin{pmatrix} \mathbf{0}^{7\times 1} \\ n_\perp^2 \\ \mathbf{0}^{20\times 1} \end{pmatrix}, \begin{pmatrix} \mathbf{0}^{7\times 1} \\ n_\perp^3 \\ \mathbf{0}^{20\times 1} \end{pmatrix}, \\
&\begin{pmatrix} \mathbf{0}^{11\times 1} \\ 1 \\ \mathbf{0}^{19\times 1} \end{pmatrix}, \begin{pmatrix} \mathbf{0}^{12\times 1} \\ 1 \\ \mathbf{0}^{18\times 1} \end{pmatrix}, \cdots, \begin{pmatrix} \mathbf{0}^{30\times 1} \\ 1 \end{pmatrix} \Bigg\},
\end{aligned}
$$

that is, spanned by 29 linearly independent vectors.

## Appendix C: Implications for uniform-velocity configurations

In this appendix we apply the well-posedness results of the conformal BDNK system to justify that there can not be conformal fluid configurations with constant velocity and variable temperature. This is a common assumption when studying analytical fluid dynamics and, moreover, there exist physical scenarios in which in principle this could be the case. For instance, in the Blasius boundary layer [91, 92] or from idealized plug flows in chemical reactors [93, 94], a fluid may display a steady velocity through a pipe with a variable temperature profile due to heat transference. Also, in some planetary atmospheres, winds may have nearly steady velocity fields, while temperature could locally vary due to solar heating or radiative cooling, among other effects. In all of the previous examples, though, a matter density is needed. For conformally invariant fluids, instead, this condition may impose strong constraints on the physical behavior of the fluid, as temperature and velocity are tied by conformal symmetry. Thus, one would expect that a configuration like this is not generally allowed except, perhaps, for very particular (fine tuned) temperature configurations. Here we show that, as a consequence of the well-posedness of the BDNK conformal theory, as well as of the correct propagation of the constraints, this cannot be possible.

In order to see so, let us assume a uniform-velocity configuration in flat space; i.e., such that $\nabla_a u^b = 0$. In this case, the full BDNK system (7)-(12) reduces to the following subset:

$$
\begin{aligned}
\dot{A} + \nabla_a Q^a + \frac{4\epsilon A}{3\chi} &= 0\,, && \text{(C1)}\\
\dot{Q}^a + \frac{1}{3}\Pi^{ab}\nabla_b A + \frac{4\epsilon}{3\lambda} Q^a &= 0\,, && \text{(C2)}\\
\dot{Q}^a - \frac{\lambda}{3\chi}\Pi^{ab}\nabla_b A &= 0\,, && \text{(C3)}\\
\dot{X} - \frac{A}{3\chi} &= 0\,, && \text{(C4)}\\
\lambda\Pi^{ab}\nabla_b X - Q^a &= 0\,, && \text{(C5)}
\end{aligned}
$$

where the dot on top of a variable indicates derivative along $u^a$; i.e., $\dot{F} \equiv u^a \nabla_a F$; $X \equiv \log(\theta)$. Due to conformal invariance requirements, the coefficients $\chi$ and $\lambda$ depend on $\theta^3$; i.e, $\chi = \chi_0 \theta^3$, $\lambda = \lambda_0 \theta^3$, with $\chi_0, \lambda_0 > 0$. Although the condition $\nabla_a u^b = 0$ implies that $S^a = 0$ and ${S^a}_b = 0$, two issues should be pointed out here. Firstly, there are two equations for $\dot{Q}^a$, for which one of them should be used as evolution equation for $Q^a$ and, as a consequence, the other one becomes a differential constraint. Secondly, there is another differential constraint coming from the definition of heat flux; i.e., given by Eq. (C5). This is certainly a differential constraint, as it only involves spatial derivatives of the evolution fields, driven by the remaining evolution equations.

Let us first inspect the propagation of (C5). By defining the quantity

$$C_1^a \equiv Q^a - \lambda \Pi^{ab} \nabla_b X, \tag{C6}$$

one should be able to prove that, if $C_1^a = 0$ at $t = 0$, then, *as a consequence of the evolution equations*, it must be $C_1^a \equiv 0$ for all subsequent times; that is, $C_1^a$ should consistently propagate. As it was done in Section III, in order to prove constraint propagation one should show that it satisfies an *homogeneous* equation, typically of the form

$$\dot{C}^a + M^{ab} C_b = 0\,, \tag{C7}$$

and then proving that (C7) admits a *well-posed* initial-value problem (for instance, by showing that it is strongly-hyperbolic). Then, by uniqueness, setting $C^a = 0$ as initial data would automatically imply that $C^a \equiv 0$ along the whole evolution. Indeed, taking a derivative of Eq. (C6) along $u^a$, we get

$$\begin{aligned}
\dot{C}_1^a &= \dot{Q}^a - \lambda \Pi^{ab} \nabla_b \dot{X} - \dot{\lambda} \Pi^{ab} \nabla_b X \\
&= \dot{Q}^a - \frac{\lambda}{3\chi} \Pi^{ab} \nabla_b A - \frac{\lambda A}{3} \Pi^{ab} \nabla_b \chi^{-1} - 3\lambda \dot{X} \Pi^{ab} \nabla_b X \\
&= \lambda \left( \frac{A}{\chi} - 3\dot{X} \right) \Pi^{ab} \nabla_b X \\
&= 0\,,
\end{aligned} \tag{C8}$$

where in the second line we have used Eq. (C4) and that $\lambda \propto \theta^3$. Then, by choosing Eq. (C3) as the *evolution equation* for $Q^a$, we get that the first two terms of the second line get canceled and, finally, by using that $\chi \propto \theta^3$ and Eq. (C4) once again, we get $\dot{C}_1^a = 0$. Thus, $C_1^a$ is constant along the integral lines of $u^a$, assuring a correct propagation.

Now, as Eq. (C3) has been necessarily chosen for evolving $Q^a$, by replacing it into Eq. (C2) we get a second differential constraint, namely

$$C_2^a \equiv \frac{4\epsilon}{\lambda} Q^a + \left( 1 + \frac{\lambda}{\chi} \right) \Pi^{ab} \nabla_b A\,.$$

Let us assume, by contradiction, that $C_2^a$ is also conserved as a consequence of equations (C1), (C3) and (C4). Then, we have $C_2^a \equiv 0$ everywhere, which also means that $\nabla_a C_2^a \equiv 0$. By direct calculation, the latter condition implies the following equation for $X$:

$$\begin{aligned}
&3\chi \ddot{X} + 4\epsilon \dot{X} + 9\chi \dot{X}^2 + \lambda \Pi^{ab} \nabla_a X \nabla_b X \\
&- \frac{3\lambda\chi}{4\epsilon\theta} \left( 1 + \frac{\lambda}{\chi} \right) \Big[ \Delta \dot{X} + 3\dot{X} \Delta X \\
&+ 6 \nabla^a X \nabla_a \dot{X} + 9\dot{X} \Pi^{ab} \nabla_a X \nabla_b X \Big] = 0\,,
\end{aligned} \tag{C9}$$

where $\Delta$ stands for the Laplace operator in flat space, and we have used Eqs. (C1) and (C4) to express the divergence of $Q^a$ as a function of $X$ and $\epsilon$, namely

$$\nabla_a Q^a = -3\chi \ddot{X} - 4\epsilon \dot{X} - 9\chi \dot{X}^2.$$

But taking now a derivative of Eq. (C4) along $u^a$, we get that, as a consequence of the evolution equations, $X$ must also satisfy the equation

$$3\chi \ddot{X} + 4\epsilon \dot{X} + 9\chi \dot{X}^2 + \lambda \Delta X = 0, \tag{C10}$$

where we have used Eqs. (C1) and (C5), but we have not used the propagation of $C_2^a$. Comparing Eqs. (C9) and (C10), it is straightforward to see, by direct comparison, that a solution of (C10) is *not* necessarily a solution of (C9) given the same initial data. This implies that for any solution of (C10), it holds $\nabla_a C_2^a \neq 0$, contradicting the initial assumption of uniform-velocity configurations.

### Appendix D: Evolution system for $\{A, Q^1, S^1, S^{11}\}$

In this Appendix we write down the system of four evolution equations obtained from the covariant set (7)-(12), using the lower-order functions explicitly given in Appendix A. The following is a $4 \times 4$ linear system for the time derivatives $\{\partial_t A,\, \partial_t Q^1,\, \partial_t S^1,\, \partial_t S^{11}\}$, whose solution gives the explicit evolution equations for $\{A, Q^1, S^1, S^{11}\}$.

In order to get this system, we have already eliminated the terms containing $\partial_t u^1$ and $\partial_t \theta$ using equations (32), as their evolution equations are decoupled from the rest of the system.

The equations are the following:

From Eq. (7), we get

$$\sqrt{1+(u^1)^2}\,\partial_t A + \frac{u^1}{\sqrt{1+(u^1)^2}}\,\partial_t Q^1 + u^1\partial_x A + \partial_x Q^1 - \frac{2Q^1u^1}{3+5(u^1)^2+2(u^1)^4}\,\partial_x u^1 - \frac{3Q^1}{\theta(3+5(u^1)^2+2(u^1)^4)}\partial_x\theta$$
$$+\frac{3(Q^1)^2}{\lambda(1+(u^1)^2)(3+2(u^1)^2)} + \frac{Q^1(\chi S^1(3+2(u^1)^2)) - Au^1}{\chi(1+(u^1)^2)(3+2(u^1)^2)} - \frac{4(\eta\chi(S^{11})^2 - A(1+(u^1)^2)(\chi S^{11}+\varepsilon\theta^4(1+(u^1)^2))}{3\chi(1+(u^1)^2)^2} = 0\,;$$

from Eq. (8), we have

$$u^1\sqrt{1+(u^1)^2}\,\partial_t A + 3\sqrt{1+(u^1)^2}\,\partial_t Q^1 - \frac{4\eta u^1}{\sqrt{1+(u^1)^2}}\partial_t S^{11} + (1+(u^1)^2)\,\partial_x A + 3u^1\,\partial_x Q^1 - 4\eta\partial_x S^{11}$$
$$+\frac{8\eta S^{11}u^1}{3+5(u^1)^2+2(u^1)^4}\,\partial_x u^1 + \frac{12\eta S^{11}}{\theta(3+5(u^1)^2+2(u^1)^4)}\partial_x\theta + \frac{4\eta S^{11}(S^{11}u^1+3S^1(1+(u^1)^2))}{(1+(u^1)^2)^2} + \frac{4\eta A S^{11}u^1}{\chi(3+5(u^1)^2+2(u^1)^4)}$$
$$+\frac{Q^1}{\lambda}\left(4\varepsilon\theta^4 + \frac{24\eta S^{11}}{3+2(u^1)^2}\right) + 4AS^1 - \frac{3Q^1}{\lambda(1+(u^1)^2)}\left((8\eta-2\lambda)S^{11}+\lambda S^1u^1\right) = 0\,;$$

also, from Eq. (9), we get

$$-\frac{1}{\chi}u^1\sqrt{1+(u^1)^2}\,\partial_t A + \frac{3}{\lambda}\sqrt{1+(u^1)^2}\,\partial_t Q^1 - 3\sqrt{1+(u^1)^2}\,\partial_t S^1 + \frac{u^1}{\sqrt{(1+(u^1)^2)}}\,\partial_t S^{11} - \frac{1+(u^1)^2}{\chi}\,\partial_x A + \frac{3u^1}{\lambda}\partial_x Q^1$$
$$-3u^1\partial_x S^1 + \partial_x S^{11} - \frac{6S^{11}u^1}{3+5(u^1)^2+2(u^1)^4}\,\partial_x u^1 + \frac{6S^{11}(u^1)^2}{\theta(3+5(u^1)^2+2(u^1)^4)}\partial_x\theta + \frac{9Q^1S^{11}}{\lambda(1+(u^1)^2)(3+2(u^1)^2)}$$
$$-\frac{S^1(2\lambda S^{11}+3(Q^1-\lambda S^1)u^1)}{\lambda(1+(u^1)^2)} - \frac{A(-2S^{11}(u^1)^3+S^1(3+5(u^1)^2+2(u^1)^4))}{\chi(1+(u^1)^2)(3+2(u^1)^2)} = 0\,;$$

and finally, from Eq. (10), we obtain

$$-u^1\sqrt{1+(u^1)^2}\,\partial_t S^1 + \sqrt{1+(u^1)^2}\,\partial_t S^{11} - (1+(u^1)^2)\,\partial_x S^1 + u^1\,\partial_x S^{11} - (S^1)^2 + \frac{S^{11}(S^{11}-S^1u^1)}{1+(u^1)^2} = 0\,.$$